# Implementation Guidance for Carbon Sequestration via Wood Harvesting and Storage

Version 1.0

Ning Zeng, University of Maryland
Daniel Sanchez, University of California - Berkeley
Erica Belmont, University of Wyoming
Henry Hausmann, University of Maryland

August 22, 2023

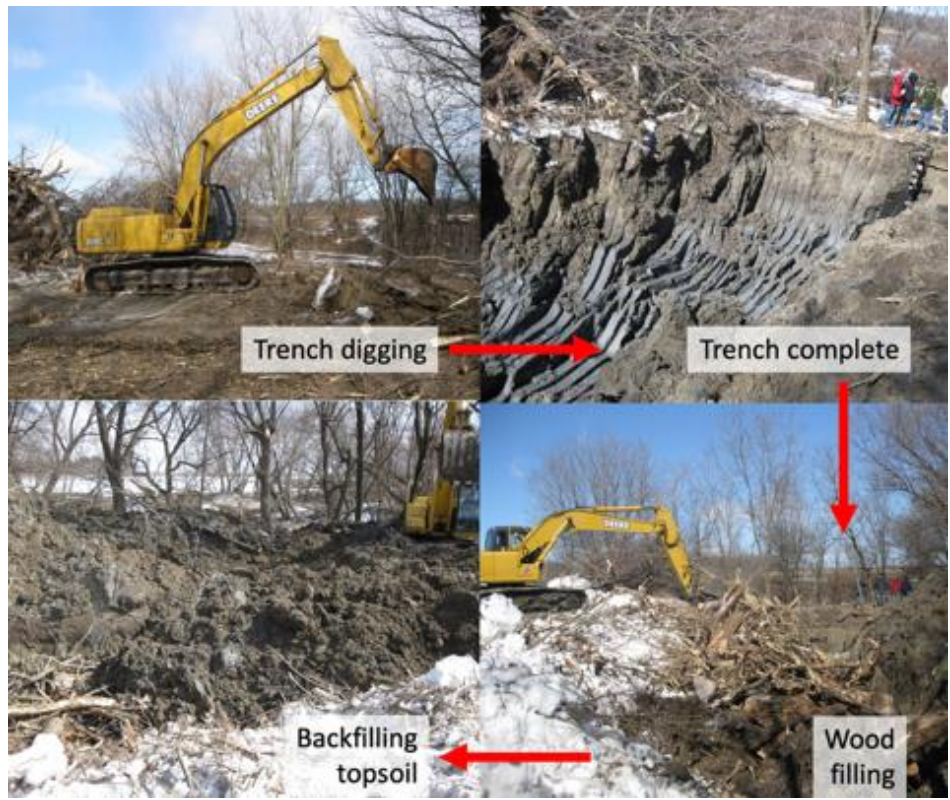

# Acknowledgements

This document greatly benefitted from reviews provided by the following individuals:

- Toby Bryce, Carbon Lockdown
- Sinead Crotty, Yale Carbon Containment Lab
- Justin Freiberg, Yale Carbon Containment Lab
- Zeke Hausfather, Frontier Climate
- Robert Hoglund, Marginal Carbon AB
- Kimberly Mayfield, Lawrence Livermore National Laboratory
- Peter Minor, Carbon180
- Charlotte Levy, Carbon180
- Anastasia O'Rourke, Yale Carbon Containment Lab
- Jimmy Voorhis, Kodama Systems
- Christopher Williams, Clark University

We also thank three anonymous reviewers from a previous version of this document, as well as David Tomberlin (National Oceanic and Atmospheric Administration) and John Lin (Exaquest Carbon) for organizing that review process.

**Cover Photos**: Main steps of a Wood Vault construction and burial during the Montreal Project where 35 ton of waste woody biomass was buried at a site near Montreal in the St. Lawrence River Valley, March 22, 2013. Clockwise from upper left: trench digging, trench complete, wood filling, backfilling topsoil. Photos by Ning Zeng.

# Preface

Wood harvesting and storage (WHS) is a promising, novel utilization for woody biomass residuals – byproducts of forest management which would otherwise decay or burn – which can contribute to a growing need for atmospheric carbon dioxide removal (CDR). During a WHS project, woody biomass is collected, transported to a storage site, and stored in a shallow geological layer in terrestrial settings where the rate of decomposition of the material is drastically slowed. Such a storage structure is called a Wood Vault.

Among biomass-based CDR pathways, WHS occupies a unique niche. Whereas other utilizations produce co-products (i.e. energy, agricultural amendments, etc.), the purpose of WHS is solely CDR. If the carbon storage is accurately quantified, monitored, and validated, the product of WHS is CDR credits. Three qualities make WHS a compelling CDR pathway:

- **Low cost:** The overall cost of WHS CDR depends on the specific storage method, availability of suitable construction materials, transport distance, and labor. However, the simplicity of the approach means low storage costs are feasible. Three sources have independently arrived at potential costs <$70/metric ton $CO_2e$[1], well below the DOE Carbon Negative Shot cost target of $100/metric ton[2].

- **Small, low-impact infrastructure requirements:** WHS sites do not require extensive, capital-intensive infrastructure. Rather, storage sites can be prepared with widely available construction equipment, natural materials, and temporary structures (e.g. trailers). After construction, sites can be restored in congruence with the local environment.

- **Distribution enables short transport distances:** Wood wastes are heavy, bulky, and often wet, presenting transportation challenges when aggregating biomass at centralized facilities. Unlike other forms of biomass-based CDR which require co-location with wood product end-markets and $CO_2$ injection sites, WHS sites can be scaled to feedstock supply volumes, distributed across the landscape, and located close to feedstock sources. This enables community-scale WHS storage sites to utilize waste biomass which may otherwise have no feasible market.

Because of these qualities, WHS has the potential to contribute significantly to near-term biomass utilization and CDR deployment. Given the urgency of the climate and wildfire crises and the nexus of WHS to forest management objectives[3] and CDR objectives[2] in the United States, WHS could benefit from investment in governance and research and development, including:

- **Development of a regulatory pathway for WHS**: In the United States, the indefinite underground storage or land application of materials are regulated for the protection of water resources, typically under the authority of the U.S. Environmental Protection Agency and cooperating state agencies. WHS facilities do not neatly fit into existing regulatory pathways. Regulation of WHS facilities should establish reasonable, performance-based criteria for water resources protection considering the biogeochemical character of natural wood.

- **Deployment of pilot-scale projects under accelerated permitting processes**: Policymakers and project developers should accelerate pilot-scale (1,000 to 10,000 metric ton) testing of WHS across feedstocks, storage types, and climate zones. Such sites typically have small footprints, less than 1 acre.

- **Research and development**: The durability of WHS depends on the level to which activity of decomposing biota can be eliminated or reduced. Continued research is needed on the mechanisms of wood decay and its control (via storage site selection, Wood Vault design, bioengineering, wood treatments, etc.), minimization of potential negative impacts, monitoring and verification, biomass availability, and suitable soil distribution.

---

[1] The three sources are Wood Vault for carbon removal (Zeng and Hausmann, 2022), Frontier CDR Purchase Application (Kodama Systems and Yale Carbon Containment Lab, 2022), and analysis of loblolly pine WHS from Scalable, economical, and stable sequestration of agricultural fixed carbon (Yablonovitch and Deckman, 2023).
[2] US Department of Energy Carbon Negative Shot.
[3] USDA Forest Service Wildfire Crisis Strategy.

This implementation guidance document is meant to inform the development of WHS carbon accounting (including baseline and counterfactual analysis), wood sourcing, Wood Vault construction, and monitoring and verification. It has been reviewed by industry stakeholders, policymakers, and scientists working on WHS and other CDR technologies. Others may use it freely for the science and practice of WHS.

Sincerely,

Jimmy Voorhis, Kodama Systems
Head of Biomass Utilization and Policy

Daniel Sanchez, University of California, Berkeley
Professor of Cooperative Extension

Ning Zeng, University of Maryland
Professor

Toby Bryce, Carbon Lockdown
Head of Commercialization

# Executive Summary

This implementation guidance focuses on carbon removal and sequestration via wood harvesting and storage (WHS), a process where woody biomass, with the embedded carbon, is stored for long timescales in shallow geologic storage. The engineering structure designed to ensure such durable storage by preventing biomass decomposition is called a Wood Vault.

This guidance contains the requirements for a basic Wood Vault project, and is intended to aid project developers, verifiers, and registries in this space. It describes a set of requirements that govern the end-to-end process of carbon removal and sequestration. This includes carbon accounting, wood sourcing via wood residual (WR) utilization, Wood Vault construction and maintenance, as well as processes for monitoring, verification, and credit issuance. Carbon accounting requirements include baseline, or counterfactual specification, and full life cycle analysis (LCA) within a specified process boundary. For the vault itself, the guidance describes a buried vault with the burial chamber covered by a layer of low permeability material to create anoxic condition. Other types of vaults can also be used to adjust to local environmental, transport, and economic constraints. Monitoring and verification requirements include in-situ sensors, gas sampling, sample excavations, and site maintenance. This guidance also contemplates land ownership and legal assurances, as well as environmental and societal impact assessments.

The implementation guidance concludes with recommendations regarding auditing, certification and carbon credit issuance.

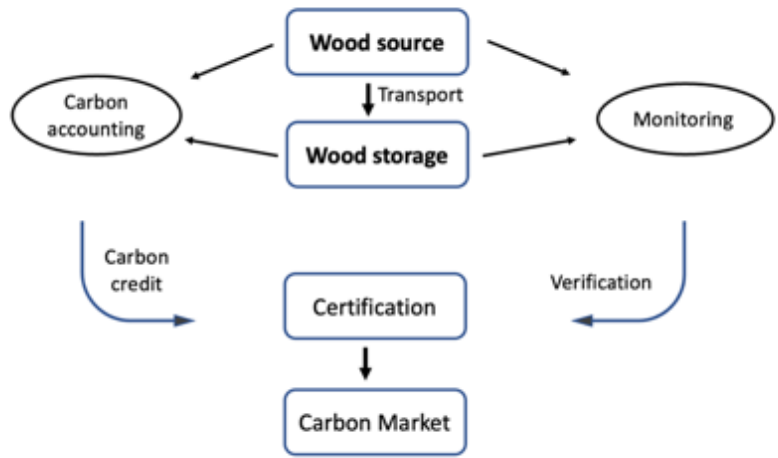

## Table of Contents

# 1 Introduction

This implementation guidance describes how to construct, monitor, and verify a Wood Vault project wherein Wood Harvesting and Storage (WHS) is pursued for purposes of carbon removal and storage. We present carbon accounting, project implementation, monitoring and verification, as well as credit issuance and certification requirements. The scientific background of Wood Vault as a technique for biomass-based Carbon Removal and Storage (BiCRS) (Sandalow, et al. 2021) is described in the Chapter 6.

A Wood Vault is a type of carbon storage facility. A Wood Vault collects woody biomass including a variety of wood residuals such as waste wood, wood from storm or fire damage, wood from forest thinning. The burial chamber is constructed to ensure anoxic conditions to prevent wood decomposition, thus storing the embedded carbon permanently (Fig. 1.1). Because the chamber can be designed to be anoxic, studies indicate that buried carbon can remain stored for longer than 1000 years (Zeng and Hausmann, 2022). Similarly, because WHS is a deliberate intervention in natural systems to avoid atmospheric return, Wood Vaults are by their nature physically additional, meaning that carbon removal would not have occurred without the intervention.

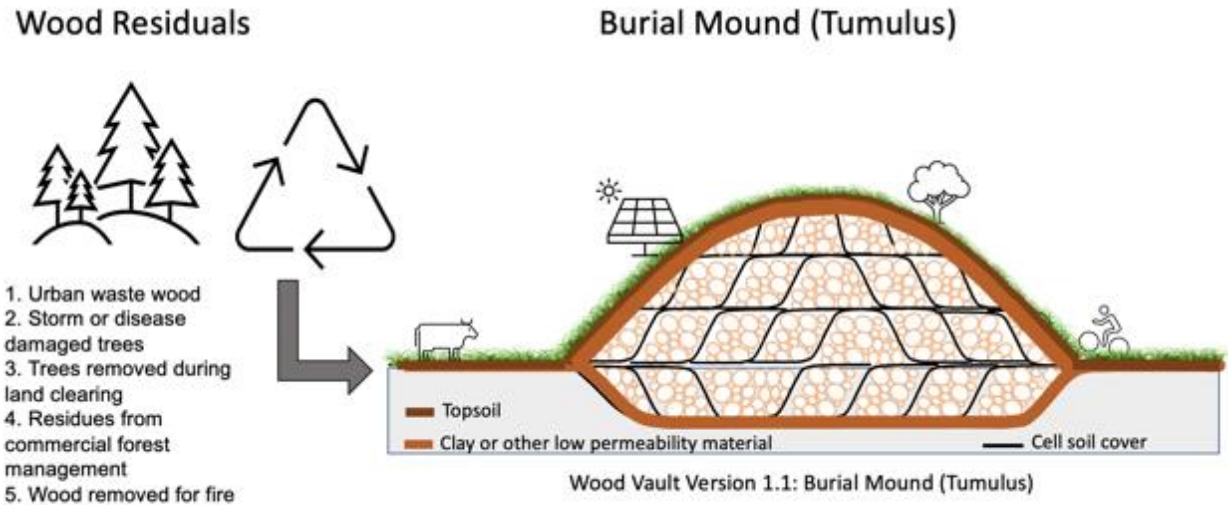

*Figure 1.1. Wood Vault is a type of carbon storage facility that sources wood sustainably, buries it in anaerobic or other condition to prevent decomposition, and stores the embedded carbon for long durations. This diagram shows one such design of a wood vault, though other designs also exist.*

Typically, the material would otherwise be released to the atmosphere as a greenhouse gas. Storing the material in a Wood Vault would directly mitigate those emissions.

Because the stored carbon was recently removed from the atmosphere by photosynthesis, not from the geological past, the method is considered carbon removal rather than avoided emissions.

Chapter 2 describes the basics of carbon accounting, including details that would be necessary for a life cycle analysis (LCA) of a specific Wood Vault project. It begins with definitions of the project boundaries and baseline. It follows with a detailed guideline on how to account carbon for a particular project. The chapter also discusses the long-term durability of sequestered carbon.

Chapter 3 is a step-by-step guide on what is required to implement a Wood Vault project. It begins with details on sourcing woody biomass. Then it outlines actual vault construction, including notes on how to modify the basic Wood Vault design into an ensemble of versions designed to meet specific needs and conditions. The chapter also includes details on how to increase durability of sequestered woody biomass, both with legal frameworks and maintenance guidelines.

Chapter 4 describes the details and what steps are necessary for monitoring and verifying the long-term sequestration of stored woody biomass, as well as the sustainability of woody biomass sourcing and acquisition.

Chapter 5 closes the implementation guidance with an explanation of the credit issuance process. It lists what data and documentation are required for creating carbon credits.

Finally, Chapter 6 contains the scientific background of Wood Vaults that supports and enhances the rest of the implementation guidance.

# 2  Carbon Accounting Requirements

Strict carbon accounting is essential to ensure high-quality projects. The process flow diagram shown (Fig. 2.1) lays out the primary operations involved in Wood Vaults and sets the system boundary for life cycle analysis (LCA), including the key aspects of Wood Vault construction and operation: the supply chain, system boundary, and baseline uses of biomass.

- **Collection and transport of biomass from source to burial site**. Process boundary in this guide does not include source forest management because this guide covers only the use of wood residuals from other operations outside the project boundary.
- **Storage of the biomass on site temporarily** before burial in order to accumulate enough material. This may not be needed if sufficient quantities are available at once.
- **Construction and maintenance of the vault,** operations to bury the wood, and ongoing monitoring and maintenance.

Process Flow Diagram of Wood Vault

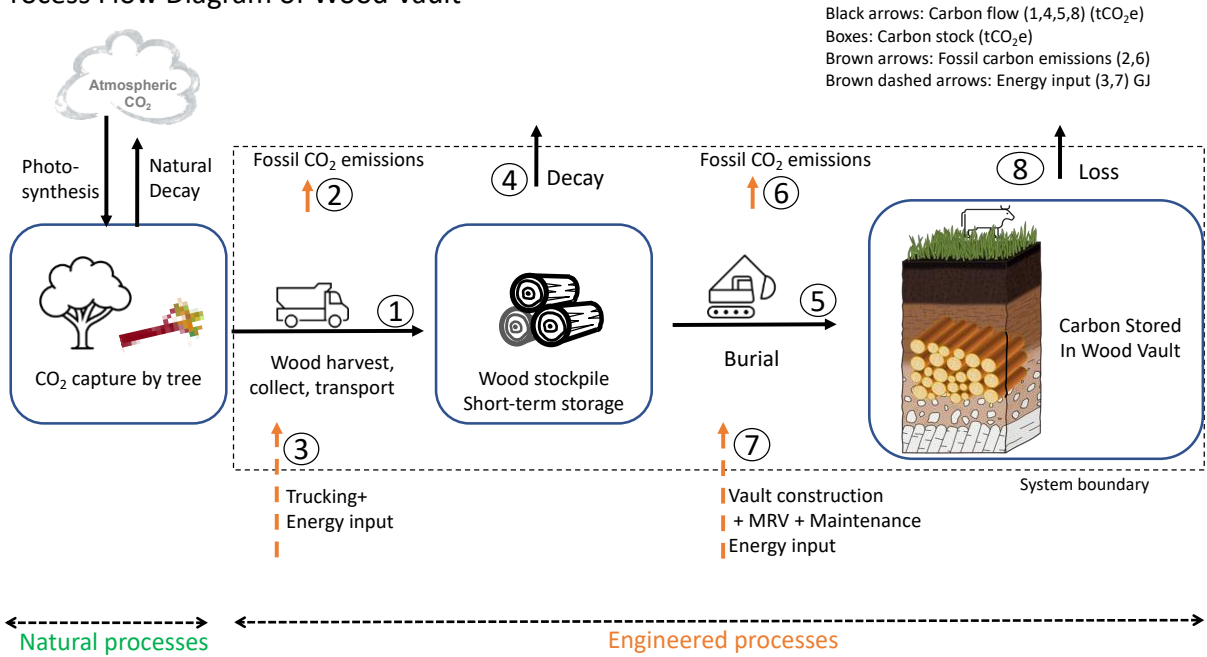

*Figure 2.1. Process flow diagram (PFD) of a Wood Vault, including both biologically derived carbon (black arrows), fossil fuel $CO_2$ emissions (solid brown arrows) and energy flow (dashed brown arrows) from machine operation. Process boundary here starts from collection and transport of wood residuals, and ends at durably stored carbon in the vault.*

The net carbon sequestration (NCS) of the full life cycle process is:

$$NCS = C_{init} - C_{Loss} - C_{Emis} - C_{LU} - C_{Baseline} \qquad (2.1)$$

Where $C_{init}$ is the volume of carbon initially stored in the Wood Vault, $C_{Loss}$ is the loss of carbon stored over time (Chapter 2.2), $C_{Emis}$ is carbon emissions of machine operation (Chapter 2.3), $C_{LU}$ is the carbon loss from land-use change and soil disturbance at the burial site and associated facilities (Chapter 2.4), $C_{Baseline}$ is the baseline carbon storage (Chapter 2.5). Note that many of these terms are time dependent including loss, land use and baseline.

## 2.1 Process boundary for carbon accounting

The process boundary is used for tracking carbon for life cycle analysis purposes. Carbon that enters the project boundary is considered stored (sequestered) while carbon exiting the process boundary is considered fugitive (lost or returned). As such, it is a key factor when determining the scale, efficiency, and success of a sequestration project.

The process boundary depicted in Figure 2.1 shows how the hybrid biological-engineering approach of WHS takes advantage of photosynthesis byproducts, followed by engineered storage. This guide addresses only the use of wood residuals so that the process boundary starts from wood collection. The NCS depends on wood sourcing, wood collection and wood vault construction & operation. The LCA must therefore account for all carbon emissions within this boundary.

In a standard project the process boundary is the project site, storage or holding site, burial chamber, and transport vectors from wood residual creation to the site. Sequestration occurs upon burial and is discounted by direct greenhouse gas (GHG) emissions from the project site as well as carbon released from biomass acquisition and transport. Project managers must provide all the carbon and energy flow and the raw data depicted in Fig. 2.1.

## 2.2 Durability and permanence of stored carbon

### 2.2.1 Determination of durability and permanence of stored carbon

A well-constructed and monitored Wood Vault should attain a 1000+ year lifetime for the main lignocellulose component of wood barring external disturbances (see Chapter 6). Since uncertainty remains regarding the impact of design and management practices on the durability of stored wood, this implementation guidance provides a framework for developing quantitative measures for durability.

**Step 1**. Because the key to a successful Wood Vault is to create anaerobic conditions that prevent decomposition, carbon durability critically depends on the low permeability seals to ensure atmospheric oxygen does not enter the burial chamber. For 1000+ year durability, a minimum thickness of 1 meter of material with permeability less than $10^{-8}$ m/s (as measured by saturation hydraulic conductivity) or equivalent is required. This defines a base decomposition rate which is the inverse of the decay time scale.

**Step 2**. The full decomposition rate is estimated with the following general formula, which includes moisture, temperature, oxygen level, and surface area (the main factors influencing the rate of decomposition of stored wood).

$$D(w, T, O, A) \ = \ D_0 * D_1(w) * D_2(T) * D_3(O) * D_4(A) \qquad (2.2)$$

where $D_0$ is the base decomposition rate which may depend on species and other factors, while $D_1$-$D_4$ represent the effect of moisture $w$, temperature $T$, oxygen level $O$, and total surface area of buried wood A. These functions can interact and be highly non-linear. For instance, at low moisture, decomposition rate is slow, increases at higher moisture, but greatly reduced (near-zero) at extremely high moisture, i.e., when water-logged.

**Step 3**. Once the decomposition rate is determined, the evolution of stored carbon can be predicted with a simple exponential decay model:

$$C_{store}(t) = C_{init} exp(-\frac{t}{\tau}) \qquad (2.3)$$

where $t$ is elapsed time, $\tau$ is the decay time which can be calculated as the inverse of decay rate $D$ obtained from Eq. 2.2:

$$\tau = \frac{1}{D} \qquad (2.4)$$

The lost carbon is:
$$C_{Loss} = C_{init} - C_{store} = C_{init}(1 - exp\left(-\frac{t}{\tau}\right)) \qquad (2.5)$$

Theoretically, $\tau$ is a continuous function of factors in Eq. 2.2. In practice, we recommend assessing durability at several timescales: 100, 500, 1000 and 10000 years, in consideration of the hundreds of years needed for a meaningful carbon sink and climate impact. For instance, Archer et al. (1997) estimated multiple timescales for neutralization of fossil fuel emitted to the atmosphere, and found a neutralization time scale $\tau_{neutralization}$ at the value of 365y for ocean $CO_2$ uptake, 5,500 y for sea floor $CaCO_3$ neutralization, 8,200 y for terrestrial $CaCO_3$ neutralization, and 200,000 y for silicate weathering. However, a commonly used neutralization timescale for carbon sequestration target is $\tau_{neutralization}$ =100 years, based on a balanced consideration of carbon neutralization timescales and socio-economic factors, which we use to assign buffer requirements below.

The durability timescale $\tau$ must be initially estimated after careful evaluation of the vault construction material, environmental condition and engineering details. Different components of the wood have different degradability, for instance between wood extractives, or chemicals found in wood, and the main lignocellulose component. Extractives such as sap account for typically less than 10% of wood biomass, which we conservatively assume would be degraded in a few years, and is subtracted from subsequent accounting.

For the remaining lignocellulose fraction, use the 4 durability timescales specified above. Using Equation 2.3, carbon remaining ($C_{store}$) after time $t = \tau_{neutralization} = 100$ years, a commonly used climate timescale, for the 4 durability timescales is calculated in Table 2.1.

Table 2.1. Carbon remaining after 100 years for different decay timescales $\tau$. Also shown is recommended buffer pool as carbon lost plus additional 5%.

| Decay time scale $\tau$ (years) | 100 | 500 | 1,000 | 10,000 |
|---|---|---|---|---|
| Carbon remaining ($C_{store}$) after 100 years | 37% | 82% | 90% | 99% |
| Recommended buffer pool (Loss+5%) | 68% | 23% | 15% | 6% |

## 2.2.2 Buffer pool based on durability

We require sites to measure and report durability based on their ongoing MRV operations. Until durability is measured, reported, and verified, we recommend a fraction of produced credits be placed in a buffer pool. This buffer pool is the expected loss plus additional 5% of initial carbon for security. The final recommended buffer pool, as a fraction of total carbon storage in the Vault at time of burial, for different durability time scale $\tau$ is listed in Table 2.1.

We recommend an initial 68% buffer pool, corresponding a minimum $\tau$ =100y. After proof of higher durability, credits can be released after verification. For example, if verified to be $\tau$ =500y, the fraction difference of 45% (68%-23%) between the two durability scales can be released from the buffer.

In summary, the above buffer pool specification constitutes an assurance of the specified fraction should the project meet a minimum $\tau$ =100y. Credits held in the buffer pool are released only when assessed and verified as stored (Chapter 4). Future work will refine the quantitative aspects of durability as scientific advances reduce uncertainty.

## 2.2.3 Prevention of methane generation

A well-constructed Wood Vault creates an anoxic environment that prevents action of main decomposers of organic matter such as fungi and insects. This leaves only anaerobic bacteria which could potentially generate $CH_4$. However, methanogenic bacteria degrade lignin, a complex polymer that forms a protective layer in woody biomass around cellulose, at extremely low rates (Colberg 1998, O'Dwyer, Walshe, and Byrne 2018). Additionally, a Wood Vault buries only clean Coarse Wood Biomass (CWB), with relatively small amounts of nutrients to support a bacteria ecosystem. Thus, methane generation is expected to be minimal in a well-constructed Wood Vault.

There may be the exception of a small quantity of wood extractives that contain more digestible molecules such as saccharides and protein. Should this component become methane, the rate is expected to be low and mostly oxidized into $CO_2$ by methanotrophs (bacteria that consume methane) in the aerobic topsoil layer above should it slowly migrate upwards (Bay, et al. 2021). This process is further slowed because the coarse woody material physically hinders bacteria invasion into the interior (Bjordal, Daniel, and Nilsson 2000). However, before all these are fully established and quantified empirically in the real-world Wood Vault burial environment, direct measurement of potential $CH_4$ leakage or the lack of is important.

If the Wood Vault is not well constructed or sealed, a partially anaerobic condition can generate methane. If a substantial amount of nutrient-rich material such as foliage is present, $CH_4$ generation could become a concern as the foliage degrades anaerobically. In this case, methane capture can be attempted as a mitigation strategy, but the methane generated may be at an intermediate rate that is too small to be effectively captured (unlike a landfill where a large quantity of methane is generated from bacterial consumption of food scraps, even in which case leakage is common), and too large to be consumed by soil bacteria in the aerobic zone. The methane emissions can be calculated with the greenhouse effects deducted from carbon accounting. In severe cases where the leakage is so large as to render the net carbon accounting not worthwhile, the Wood Vault is considered failed. This guide requires the burial of only clean coarse woody biomass with total anaerobic sealing that is expected to have negligible leakage to the atmosphere. Monitoring procedure is required to ensure this is the case (Chapter 5).

## 2.2.4 Carbon loss caused by disturbance

Disturbance refers to an external disruption of the stored carbon, exposing it to fresh air and oxygen, or otherwise compromising sequestration. Disturbances come in two forms:

1) Human interference
2) Environmental change

In a human interference scenario, human activity compromises the sequestration of carbon, for instance, deliberate re-excavation or subterranean construction. These activities are strictly prohibited in any Wood Vault conservation easement. Preventing them is a matter of enforcement and depends on the specific conditions of the specific Wood Vault project.

Environmental change refers to the compromise of sequestration due to external environmental factors. This includes natural disasters and climate change. Natural disasters are relatively straightforward. Earthquakes, hurricanes, floods, and similar events can disrupt the structural integrity of a Wood Vault. The risks of such events can be mitigated by minimizing the exposure of a Wood Vault site to such events. Disaster propensity is a key factor to analyze in location selection.

## 2.3 Transport and construction deductions

A deduction is applied to net sequestration based on the emissions accrued during woody biomass acquisition and Wood Vault construction. The GHG emissions should be tracked from the following sources and then subtracted from net sequestration:

1) Transportation of woody biomass
2) Transportation of off-site materials such as imported clay
3) Creation of off-site materials
4) Woody biomass collection, preparation, sorting.
5) Vault excavation
6) Woody biomass burial
7) Vault sealing
8) Project related travel energy consumption and carbon footprint
9) Monitoring and verification
10) Site maintenance

From there, the sum of the fossil fuel consumption of all the above activities can be found via use of greenhouse gas emissions factors.

$$C_{emis} = \Sigma_i \, Fuel \; Consumption(i) \, \times \, Emission \; Factor(i) \qquad (2.6)$$

where $i$ represents each activity of operation. The use of emissions factors allows operators to reduce their emissions through use of electricity or biofuels.

## 2.4 Land use deduction

Land use changes can release above ground and below ground carbon back into the atmosphere. As such, any disruption in vegetation and soil carbon caused by Wood Vault construction is subtracted from the sequestration total. Before the Wood Vault construction begins, developers must conduct a survey of the carbon stocks contained in the area that will be occupied by the Wood Vault and associated facilities. From there, this value is subtracted from the sequestration total. Then, during each monitoring period, a similar survey should be conducted of the region's carbon stocks. Any changes since the last survey should be added or subtracted to the sequestration total as appropriate.

$$C_{LU} = Land \; Carbon \; (current) - Land \; Carbon \; (initial) \qquad (2.7)$$

## 2.5 Baseline

To estimate the volume and durability of the carbon sequestered, establishment of a baseline is required. This means analyzing what would have happened to all carbon within the project boundary and comparing it to what would have happened to said carbon without any intervention.

For operating wood vault burial chambers, the baseline is that the residual woody biomass would decompose and all the carbon stored within is emitted back into the atmosphere as $CO_2$ within several years (decomposition or burning). The simplest approach is to assume an exponential decay with a characteristic time scale $\tau_{baseline}$ based on the circumstance (decompose on forest floor, burned, etc.);

$$C_{baseline}(t) = C_{init} exp(-\frac{t}{\tau_{baseline}}) \qquad (2.8)$$

Suggested values of $\tau_{baseline}$ are as follows (Table 2.2).

Table 2.2. Suggested values of $\tau_{baseline}$. Estimate for decomposition based on Russel et al. 2014.

| Baseline | $\tau_{baseline}$ |
|---|---|
| Left to decompose on forest floor | 20y |
| Mulched | 5y |
| Burned | 1y |

We assess carbon storage at a minimum of 100 years (Chapter 2.2.1). Within this time frame, wood residual would have nearly all been released back into the atmosphere in most baseline scenarios. This simplifies accounting.  In the other direction, some more resistant components such as lignin and burned residue may have much longer residence time that should be accounted with more sophisticated models in the future. More precise accounting accounts for impacts of environmental variables and wood characteristics using the general form of Eq. 2.2 with specific functional forms and parameter values suitable for the wood source baseline environment.

## 2.6 Life cycle analysis

Measuring the amount of carbon captured and sequestered relies on life cycle analysis (LCA); an LCA study involves establishing baselines and conducting a thorough inventory of the energy and materials required by all the processes involved in capturing and sequestering the carbon and calculating their corresponding GHG emissions to the environment. An LCA assesses cumulative potential environmental impacts and is conducted only for processes within set system boundaries.

The PFD (Fig. 2.1) describes the processes and carbon flow that need to be accounted for. Chapters 2.1-2.5 above list the specific calculations for each factor to be considered. Follow through those steps to complete the LCA analysis, with final net carbon sequestered (NCS) using Eq. 2.1

Box 2.1 summarizes the requirements for carbon accounting.

Box 2.1. Carbon accounting requirements

The project developer must provide full carbon accounting that include:

- Clearly defined process boundary
- Quantity and durability of the stored carbon
- Carbon loss over time and $CH_4$ emissions
- Wood sourcing dependent baseline
- $CO_2$ emitted during machine operation
- $CO_2$ emission due to land use change at the burial site
- Net Carbon Sequestered on the specified durability timescale
- Additionality justification

# 3 Project Implementation Requirements

Implementation of a Wood Vault project consists of three major components: 1) the sustainable sourcing of woody biomass, 2) siting, and 3) the construction and maintenance of a Wood Vault to store carbon. Requirements and processes for project implementation are described in detail in the following two chapters.

## 3.1 Biomass sourcing

The first step of a Wood Vault project is to secure a source of woody biomass which would otherwise return to the atmosphere. There are two primary source supplies:

- Type-A: Wood residuals (WR), including urban waste wood and forestry residue.
- Type-B: Managed forests, such as a plantation or restored forest.

**This guide addresses only Type-A sourcing from wood residuals**. Type-B processes require assurance of the sustainability of harvested wood and managed forests. These will be developed in a future and distinct guide.

Some Type-A sources are commonly known as 'waste' wood (e.g., urban waste wood). Since at some future time there may be clear climate and economic value of some Type-A sources, we use the term wood residual (WR) as a general term.

### 3.1.1 Sourcing of wood residual

Type-A sourcing with wood residuals refers to the process of sourcing woody biomass from already existing supply streams that would normally result in wood residuals, including, but not limited to:

1) Urban waste wood: wood removed from yard work.
2) Storm, fire, or disease damaged trees.
3) Trees removed for land clearing.
4) Residues from commercial forest management
5) Wood removed for fire hazard reduction.

Restrictions on these sources include the following:
- Exclude wood with significant fungi colonization, which may continue decomposition after burial.

- Exclude wood that is not fully intact or that is already partially decomposed (USFS decay class 1 as defined in Russell et al. 2014 allowed– all other decay classes not allowed).
- Exclude wood that is or has been infested by insects such as beetles.
- Exclude wood that has a stem diameter less than 10 cm, and that has a long-axis length of less than 1 meter.
- Exclude phytoremediation biomass grown on brown field and mining recovery sites.
- Exclude wood sprayed with herbicides or insecticides during management and collection.
- Exclude biomass grown on lead contaminated soil from urban sites.

Two additional streams are not recommended at present if they are destined for well-managed landfills, based on their baseline:

- Construction and demolition debris
- Old furniture and other wood products

As before, consideration of baseline use of wood is paramount. Attestation and documentation could include letters from professional foresters, cooperative extension advisors or faculty, or local environmental regulators responsible for waste management within the jurisdiction.

Once the source of residual wood has been identified and contracted, the next step is to redirect it to the site of the Wood Vault. Transportation using fossil fuels will reduce the total credit volume of the project. It is critical to keep an accurate account of the carbon emitted in acquisition and transport.

Once the woody biomass arrives at the facility, it must be inspected for purity of composition. Only clean woody biomass can be buried without contaminants such as food scraps, toxic chemicals and heavy metals. Rocks, sand and gravel, non-organic soil, non-reactive cement and waste concrete are allowed provided evidence of no adverse effect on the stored wood but cannot be counted towards removed carbon.

Wood residuals should be sorted into Coarse Woody Biomass (CWB), i.e., trunks and branches with a minimum diameter of 10cm or another widely accepted value, and Fine Wood Biomass (FWB), including small twigs, woodchips, leaves and other finer material. Only CWB should be buried. FWB, which is both more subject to decomposition and nutrient rich, should be left on the surface biosphere.

## 3.1.2 Carbon measurement

The next step is to measure the sorted CWB and calculate carbon content using direct weight measurements. Calculation based on volume is optional, providing useful information for stock management and vault construction. Follow these measurements with independent, third-party verification and reconciliation of any inconsistencies among multiple methods.

**Wood Weight Method** is required for carbon measurement and calculation**, conducted as follows:**

Use the following formula to obtain CO2e ($CO_2$ equivalent of carbon stored in wood) using the wood weight, water content, and dry-basis carbon content, multiplied by a factor of 44/12 (the molecular weight ratio of $CO_2$ and C).

$$CO2e = Weight\ of\ wood \times (1 - Water\ content) \times Carbon\ content \times \frac{44}{12} \qquad (3.1)$$

Variations in water content should be measured with calibrated instruments (e.g. moisture meters) and lab analysis (e.g. wet weight and dry weight) before summing up for CO2e in variable stockpiles. Carbon content is typically 47-51% of dry matter and varies little among species (Thomas and Martin, 2012). When there is a lack of information, use the most conservative parameter values to estimate.

Box 3.1 summarizes the data requirements for wood residual sourcing.

---

**Box 3.1. Data requirements on wood sourcing from wood residuals**

The project developer must provide a report with supporting evidence on the full information of the wood source including:

- **Baseline and additionality**: The original destination and usage before the sourced wood was diverted to a Wood Vault facility, for instance, urban waste wood that would otherwise go to mulch facility for compost, or forest residue that would have been burnt or left to decompose on the forest floor. This information will be used as the baseline in carbon accounting.
- **Amount of carbon** to be stored in the Wood Vault, calculated based on weight method.
- **Fuel consumption data**: fuel type and quantity of transportation of wood, operation in receiving/building wood stockpile, separating wood types. The data will be used to calculate carbon emissions during operation.

## 3.2 Wood Vault construction

Several versions of Wood Vault are viable (Zeng and Hausmann, 2022). This guide describes two major designs and one variation.

Version 1: Burial mounds, in which woody biomass is buried in an anaerobic environment below the active soil layer. This version consists of two sub versions:

- Version 1.1: Tumulus, in which the burial chamber is partially above and partially below the original land surface ground level.
- Version 1.2: Barrow, in which the burial chamber is entirely above the ground level.

Version 2: Pit or Subterranean/Underground Vault, in which the burial chamber is entirely below ground level in a dug pit or a pre-existing mine shaft or quarry.

### 3.2.1 Basic vault construction: Tumulus

The main goal of a Wood Vault is to prevent woody biomass decomposition. For the template Wood Vault described in this chapter, the design objective is to create a fully anaerobic burial chamber by sealing the buried wood below the biologically active soil layer with minimal air or water movement. The detailed steps are described below with Fig.1.1 illustrating a completed Wood Vault (Version 1.1: Tumulus). These steps are based on current knowledge and should be modified or expanded in practice. In particular, variations may be allowed based on site condition for the best interest of preservation.

1) **Locate a suitable site** with size commensurate to the estimated woody biomass source and planned operation time horizon of the facility. Conduct a survey to assess site characteristics, including soil type, soil depth, soil profile, topography, hydrology, climate, native vegetation, environment, cost, transportation access, ownership, and local economic conditions. This survey is critical to making key engineering decisions and deciding what standards are applicable. Avoid competing with critical land needs such as housing and agriculture and give preference for using marginal soils. Other important considerations include erosion risk, height above sea level, earthquake risk, and flooding risk.

2) **Collect woody biomass** before burial following the procedures described in Chapter 2.1.

3) **Separate the stockpiled material** as described above into CWB and FWB. Remove any contaminated material. Bury CWB only. Specify the packing density (volume of wood relative to total space). Mixing FWB or other contaminants with CWB may negatively impact the durability of coarse wood. Do not bury leaves as they are both nutrient rich and more likely to decompose.

4) **Take sufficient samples of** biomass of each of the main types present, to account for the natural variability in species, size, and decay state, and preserve in cool and dry lab conditions. This will serve as a baseline for later MRV efforts. Assess each sample for moisture content, % carbon, % nitrogen, and lignin content.

5) **Excavate soil to form a large trench/pit** using a method that is fuel and resource efficient. Excavate the organic containing topsoil first and keep it separate from the subsoil. Take samples of soil at multiple depths and quantify soil carbon at each depth. Later, the organic containing topsoil will be replaced on the Wood Vault surface in order to minimize environmental impacts with an additional benefit of supporting vegetation regrowth. It is important to keep this time period brief. The woody biomass should then be buried in the subsoil (B-horizon) or lower horizon, completely isolated from the biologically active topsoil above (see item 8 below on depth and permeability specification).

6) **Divide the Wood Vault** into multiple sections (cells) for large vaults. A cell is a subdivision of a Wood Vault, separated from other cells and individually capped. Wood Vaults have cells because 1) it takes time to build enough wood stockpile, 2) the pile on the surface should not be left for too long due to slow decay, using cells with capping strikes a balance between operation efficiency and minimizing decay before enclosure of the whole Wood Vault. After a cell is fully filled, cover it with a layer of soil. Compact the soil and allow it to settle, filling any gaps in the woody biomass. If all the wood for the whole Wood Vault is available at once, fewer cells are needed.

7) **Take seasonal variations (cold vs. warm, or wet vs. dry) into account** while carrying out different steps of Wood Vault construction. Carrying out logging, collection, and excavation operations in cold/dry seasons where the ground is frozen minimizes soil damage and compaction. These conditions also allow better machine maneuverability.

8) **To ensure long-term wood preservation and high durability,** line the Wood Vault walls, floor, and ceiling with **low permeability soil, such as clay,** to maintain anaerobic conditions. Low permeability is defined as having a saturated hydraulic conductivity less than $10^{-8}$ m/s.

    a. Use the excavated local soil to cover the buried wood if it has low permeability. If low permeability material is not available on site, it can be sourced from somewhere else. The degree of anaerobic conditions also depends on seal thickness. Use a minimum of 1 meter of low permeability material.

    b. Where the topographic slope is significant, dig a ditch upslope to divert running water around the Wood Vault. Line the Wood Vault wall on the upslope facing side with additional low permeability soil or a synthetic material to prevent water from moving laterally through the buried woody biomass. This will minimize episodic reoxygenation of the burial environment.

    c. Keep the burial chamber either above the local water table maximum or below the water table minimum to avoid a fluctuating water-air boundary bringing in oxygen.

    d. Encourage compaction of soil over time through careful packing.

    e. **Final capping** consists of covering all woody biomass with at least one meter of low permeability material, either previously excavated soil or an externally sourced material. Compact the capping material to reduce the pore space inside the Wood Vault. Place the original topsoil back on top of the capped Wood Vault.

Alternative cover designs are permittable if they are compliant with the Resource Conservation and Recovery Act (RCRA) Subpart F of Title 40 of the U.S. Code of Federal Regulations.

9) **Some land settling over time** may be inevitable. Carefully packing the filling material of wood and soil can minimize but is unlikely to eliminate settling. After final capping of the Wood Vault, monitor the area for settling as gaps are filled in and the land altitude lowers slightly (MRV plan in Chapter 4). If wood eventually rots, even just partially, land settling will become significant, in which case the Wood Vault has failed. Make sure the surface is covered by either active utilization or shallow rooted vegetation.

## 3.2.2 Alternative Wood Vault construction

The above section 3.2.1 describes the half-above half-below ground Version 1.1 (Tumulus). This chapter describes the other two versions Barrow and Pit (Fig. 3.1). Most steps are the same as in Tumulus, the unique requirements for these two versions are described below.

Version 1.2: Barrow: the burial chamber is entirely above the ground level, which eliminates the impact of a fluctuating water table that may bring in air. The unique requirements for constructing barrow are:

- Requires importing soil.
- Line clay all around.
- Additional erosion control considerations.
- Additional ecological assessments.

Version 2: Pit or Subterranean/Underground Vault, in which the burial chamber is entirely below ground level in a dug pit or a pre-existing mine shaft or quarry. The unique requirements are:

- Soil is excavated and taken somewhere else. In the case of an existing quarry/mine, no additional excavation is needed.
- Burying the logs below the lowest water table creates a permanent waterlogged condition ideal for an anaerobic environment.

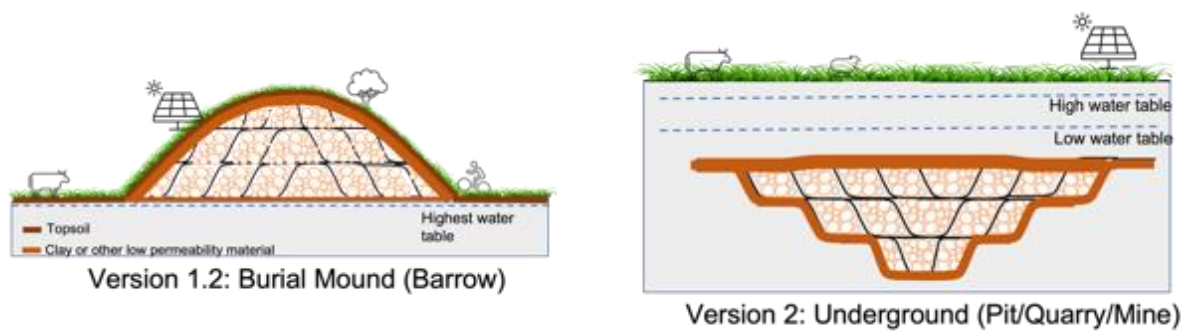

*Figure 3.1. Other versions of Wood Vault: a) Version 1.2: fully above-ground (Barrow); b) Version 2: fully underground (Pit/Quarry/Mine).*

### 3.2.3 Permitting and assessment of environmental and social impacts

Minimizing harms involves the avoidance of negative impacts to economic, social, and environmental systems stemming from a CDR project. Beyond simply avoiding harm, ideal projects will be ones that pursue co-benefits by advancing sustainable livelihoods and environmental justice, building climate resilience, supporting water conservation, and protecting ecosystems and biodiversity. These criteria are adapted from Carbon Direct and Microsoft's 2023 Update to the Criteria for High-Quality Carbon Dioxide Removal.

Project developers must:

- Show that feedstock production, biomass conversion, and carbon disposal operations have a low risk of any major negative impacts on the surrounding ecosystems (including soil health, biodiversity, water, criteria air pollution) or local communities.
● Develop and share a strategy for mitigation of adverse impacts to air, water, and land quality, including those impacts related to biomass processing and storage.
● Transparently report any use of toxic and/or persistent environmental pollutants, including agrochemicals used in the production of purpose-grown feedstock.
● Prevent community displacement when selecting facility location.

Project developers should:
• Actively promote long-term sustainable livelihoods and economic opportunities for local communities, by developing local and regional biomass-based CDR expertise.
• Explore how project activities (e.g., feedstock production and product/co-product sales) can benefit under-resourced and marginalized populations, including wealth generation and economic empowerment.

## 3.3 Wood Vault maintenance

### 3.3.1 Physical maintenance

After final settling, the surface of a Wood Vault can be used for other purposes. These include, but are not limited to, agriculture, recreation, or solar farming. Avenues must also be kept open for all MRV activities described in Chapter 4. The one maintenance requirement is to keep the layer of low permeability material surrounding the woody biomass intact. If it is breached (due to natural disaster, human interference, or other causes), repair this layer as soon as possible by either acquiring material on site or importing it from an external location. Investigate any indication that the outer layer has been breached to determine if maintenance is required.

### 3.3.2 Legal permanence

A key step before breaking ground on Wood Vault construction is to set up a legal framework for ensuring the long-term permanence of the Wood Vault. Such a legal framework is done no matter what type of Wood Vault is constructed or what variations are made in design. This is done via a conservation easement. A conservation easement is a legal structure that limits what activities can be done on a specific plot of land, set up in the name of conservation. These restrictions are monitored and enforced by a legal entity, often a local land trust. In the case of a Wood Vault, the conservation easement prevents excavation or interference with sequestered woody biomass, unless in the service of predefined MRV tasks. These restrictions will last even if the plot of land is transferred to a new owner. Specific restrictions on surface activities can vary to suit the needs of individual projects. Once the easement is set up, the land containing the Wood Vault can be sold and developed within the guidelines of the easement as long as all required MRV tasks can be fulfilled.

Box 3.2 summarizes the requirements for Wood Vault construction and maintenance.

---

Box 3.2. Wood Vault construction and maintenance requirements

The project developer must provide a comprehensive report documenting the Wood Vault construction and burial process, including photos, videos, engineering drawings and descriptions covering:

- Physical condition of the burial site: location, size, hydrology including water table and water flow, soil profile, and initial carbon measurements
- Disturbance to the site ecology, soil and hydrology due to the operation (if excavation for Wood Vault) and how it is managed and projection for future recovery. In the case of mine remediation, document the land history and potential benefit or any negative impact.
- Buried wood: quantity, species, type, size, moisture content.
- Burial chamber condition after burial: soil condition surrounding the stored wood, permeability of soil, soil layers above the wood, hydrological condition.
- Detailed engineering drawings must indicate the full setting of the Wood Vault, its geometry, as well as the inside condition, location of wood relative to soil and sub-cells.
- An assessment of environmental and social impacts.
- A plan for future maintenance, and monitoring and verification.
- Land history, ownership, and the legal framework for ensuring long-term storage.

---

# 4  Monitoring and Verification Requirements

Carbon stored in a Wood Vault is verifiable as the carbon is contained in a solid form (woody biomass) contained in an easily accessible managed central facility. Still, ensuring that carbon is durably sequestered requires monitoring, measurement, reporting and verification (MRV). This chapter goes over several MRV techniques to be applied to Wood Vault projects.

## 4.1 Monitoring and measurement of the burial environment

When initially constructing a Wood Vault, monitoring sensors or gas well samplers must be installed along with the woody biomass. The sensors or gas samples should be used to measure the following variables inside of the burial chamber:

- $CO_2$ Concentration
- $O_2$ Concentration
- $CH_4$ Concentration
- Temperature
- Pressure
- Humidity
- pH (Initial test)

In a well-constructed Wood Vault, $CO_2$ concentrations will initially increase while $O_2$ concentrations steadily drop as the leftover $O_2$ buried in the chamber is consumed by a small number of residual decomposers. Both will level off within weeks to months, with the final $O_2$ concentration settling towards zero, killing any remaining decomposers.

Once these key variables settle into a demonstrable quasi-equilibrium with negligible methane emissions, the Wood Vault can be considered geologically stable. Monitoring should last at least several years. Once the geologically stable state is achieved, monitoring frequency can be substantially reduced (Chapter 4.2 and 4.3 below).

Any deviation from these trends means there is a problem with the Wood Vault and the sequestration may have been compromised. Determine what that problem is and redress it if possible. If it is not possible to redress, the project has failed.

There are two types of sensor systems for monitoring:
1) Periodic sampling followed by lab measurements. This typically yields results with higher accuracy.

2) Continuous data collection with high temporal resolution down to minutes. This method minimizes sampling error in the presence of fluctuations and noise, but it may risk lower accuracy.

Traditional soil and hydrological monitoring typically uses periodic sampling to reduce the cost of frequent field trips. Recent advancements enable continuous monitoring of gasses with smart low-cost sensors (Martin, et al. 2017). Some of these sensors transmit data instantly to the cloud using the Internet of Things approach (IoT). Redundant, in-situ, low-cost sensors should be supplemented by periodic sampling and lab analysis to verify the sequestration of all projects.

## 4.2 Above ground flux monitoring

To monitor potential methane generation and leakage, install instruments such as soil flux chambers to monitor flux of $CH_4$ and $CO_2$. The instruments must be mounted immediately after capping of the Wood Vault. One location should be where interior gas is most likely to accumulate and escape, such as at the top of a Tumulus dome. For comparison with baseline, instruments should be mounted both on the capped Vault and the undisturbed surrounding in order to detect the difference.

Significant $CH_4$ flux originating from inside the burial chamber over a long period of time that cannot be neutralized by soil above is an indicator of Wood Vault failure (Chapter 4.5).

## 4.3 Re-excavation and analysis of buried wood samples

Another MRV technique is to excavate and retrieve samples of buried woody biomass and analyze the status of preservation. This is considered a definitive answer to the durability question and should be conducted for all medium to large scale projects in the present state of scientific and practical knowledge. Repeat this entire process every 5-10 years for the first 50 years, and every 25 years afterwards.

For this process, excavate a representative cell or section of the Wood Vault. Make sure to keep the topsoil and low permeability capping material separate from any other earth and woody biomass excavated. Ensure that the sides and floor of the cell are not interfered with or damaged in the excavation. Collect a small amount of woody biomass from a variety of depths, then return the rest of the excavated woody biomass and earth to the burial chamber. For the baseline capping process, reseal the top of the Wood Vault with the same low permeability material used in the initial capping, potentially acquiring new material if the integrity of the initial cap is compromised. Finally, take the original topsoil and recover the surface of the Wood Vault. Once the woody biomass has been excavated, take it to a qualified lab for analysis. Measure the following parameters for samples from every recovery depth and plant species in the sample batch:

1) Visual and microscopic examination
2) Density
3) Mechanical Strength
4) Chemical composition of wood: lignin, cellulose, hemicellulose
5) Carbon Content
6) Water Content

While it is recommended to analyze all the above, the goal is to have sufficient data to show the preservation status of the buried wood. Significant loss inconsistent with the stated durability/permanence is an indicator that something has gone wrong. If it occurs, begin an investigation into whether sequestration has failed.

## 4.4 Monitoring soil settlement, Wood Vault integrity and site management

After some initial soil settlement to adjust to the pore space in the burial chamber, long-term settling is an indication that decomposition may have occurred.

Some initial soil settling is inevitable after the final capping of a Wood. The degree of settling depends on initial space and the settlement may take a few years. Follow these steps to monitor soil settling and the overall Wood Vault integrity:

1) Estimate the initial space in between the buried woody biomass, subtracting the portion filled during soil backfill.
2) After enclosure of the Wood Vault, conduct a survey of the ground surface. The survey can be done using a traditional standard survey technique or aerial/drone technology.
3) Build a 3-dimensional model of the Wood Vault, with accurate information of its geometry, location of woody material, pore space vs. wood ratio, and other relevant data.
4) Repeat the survey annually until settlement is deemed complete.
5) Compare the survey data from multiple time periods to determine whether soil has completely settled which is expected to occur within 10 years in most conditions. Cross check the data with gas monitoring of the burial chamber interior. Soil settlement that significantly exceeds the initial void space, or major settlement after 10 years is an indicator of loss of woody biomass or major hydrogeological change and can be an indicator that further investigation is required (also see Chapter 4.5 on failure).

6) During each survey, document the general integrity of the Wood Vault and the environment above and surrounding the Wood Vault, including vegetation, land use and management. Use the 3-D Wood Vault model to quantify any change. Check against the original management plan, convey and reconcile any discrepancy to the relevant parties.

Box 4.1 summarizes the requirements for monitoring, reporting and verification (MRV).

Box 4.1. Monitoring, reporting, and verification requirements

- Monitor the burial environment:
  - Gases and hydrology
  - Soil settlement
- Must show that the interior of the burial chamber has become biologically and geologically stable via directly measured data.
- Periodic sampling of buried wood followed by lab analysis to show degree of wood preservation and carbon retention.
- Periodic surficial site survey to show the Wood Vault has not been disturbed.
- Periodic survey to show that site and surrounding environment is well managed.

## 4.5 Site failure

Failure occurs when any of the following occur:

1) Monitoring: non-attainment of quasi-equilibrium after 10 years
2) Re-excavation: significant deterioration of samples found after examination
3) Significant settling: resumption of settling after period of quasi-equilibrium,
4) Methane evolution: sustained leakage of methane during monitoring
5) Other significant disruptions: Earthquakes, hurricanes, floods, and similar events that disrupt the structural integrity of the Wood Vault

Crediting for carbon storage must cease once failure occurs and cannot be re-established until remedied.

# 5 Auditing, Certification, and Issuance

We conclude with recommendations regarding auditing, certification, and carbon credit issuance. While these recommendations encompass essential elements of credit issuance and certification, we expect that others will develop additional regulations.

## 5.1 Reporting and data requirements for validation

Validation of a Wood Vault project requires the following reports and data from the project developer or manager.

1) Results of carbon accounting with full LCA analysis. (Box 2.1)
2) Documentation of wood sourcing, including baseline specification and additionality justification for wood sourcing. (Box 3.1)
3) Detailed description of Wood Vault construction and burial, including site survey, excavation, sealing material, depth, dimension, quantity/size/species of wood buried. (Box 3.2)
4) Report from environmental and societal impact assessment. (Box 3.2)
5) Long-term management plan for Wood Vault maintenance, including both physical maintenance and legal and financial documents such as proof of conservation easement and land trust. (Box 3.2)
6) Monitoring and verification plan and initial results, supplemented by analysis of continuous monitoring and periodic sampling results as they become available. (Box 4.1)

Compile all this information into a well-organized single Final Project Report (FPR) to the extent possible, with supporting documents included as appendices. This report should be amended over time with additional information such as long-term monitoring results and change in land ownership as soon as available.

## 5.2 Project Evaluation Matrix (PEM)

As a high-level outcome of the full assessment of project data and documentation, a Project Evaluation Matrix (PEM) should be produced. This summary matrix including the following indicators:

- Wood residual types.
- Net carbon sequestered (NCS): net carbon stored after subtracting operational emissions, loss and baseline, corresponding to a specific durability time scale.

- Durability/permanence: theoretically continuous, but expressed in major preservation timescale categories for practical application. Assess credits generated and credits held in buffer pool 100, 500, 1000, and 10000 years.
- Summary of environmental and societal impacts.
- Buffer pool for carbon credits.

Specific category values of these criteria are given in Table 5.1. These criteria form the basis of assigning carbon credit valuation.

*Table 5.1. Wood Vault project evaluation matrix (PEM). 1 Kt = 1000 tonnes of $CO_2$ equivalent of woody biomass.*

| Wood source (Baseline) | Type-A (Wood residual): specify sub-category (natural urban waste wood, land clear, fire damage, storm blowdown etc.) | | | |
|---|---|---|---|---|
| Net Carbon Sequestration (NCS) [select one] | Large (>100 Kt) | Medium-large (10-100 Kt) | Medium (1-10 Kt) | Small (<1 Kt) |
| Assessed Durability<br><br>% of remaining lignocellulose after 100 years | Ultra-high quality (>10,000y)<br><br>99% | High quality (1,000-10,000y)<br><br>90% | Medium-high quality (500-1,000y)<br><br>82% | Medium quality (100-500y)<br><br>37% |
| Environmental and Societal Impacts [classify significant impacts in following categories] | Beneficial | Slightly negative; can be mitigated | Moderate impact | Severe impact |
| Ratio of credits in buffer to credits produced | At 10,000 y<br><br>6% | At 1000 y<br><br>15% | At 500 y<br><br>23% | At 100 y<br><br>68% |

## 5.3 Third-party verification

For validation, an accredited third-party verifier is required to conduct an independent verification of the project, including:

1) Site visit and document review to verify the nature and quality of the Wood Vault construction.
2) Review monitoring plan, data and results.

3) Review Wood Vault management plan to ensure the physical longevity of the stored carbon.
4) Review conservation easement, land ownership, and other legal documents to ensure the buried wood is not disturbed.
5) Review the assumptions in the baseline of wood source and additionality.
6) Review the accuracy of carbon accounting and life cycle analysis.
7) Review the environmental and societal impact assessment.
8) Estimate durability of stored carbon.
9) Assess buffer pool contributions.
10) Sign off on the auditing report.

## 5.4 Reporting periods & credit structure

Setting up a carbon credit for a Wood Vault project requires determining an active credit lifetime (ACL). This is the period over which active monitoring and verification takes place. There are monitoring reports due periodically throughout the AAL. An initial report is due 2 years after final capping, corresponding to the time for the Wood Vault to stabilize (Chapter 4.1). Afterwards, the reporting period can be extended to 10 years or longer, varying depending on the specific project and credit issuance scheme. Two to three of these reports should suffice to establish long term durability of sequestered woody biomass. However, these reports can continue for as long as possible if an entity exists to create them. Long-term monitoring should also continue for natural and human disturbance under the easement requirement. Credits can be released from the buffer pool as new reports are issued.

The ACL is the period for which the project is being actively monitored and giving out carbon credits. During this period carbon credits can be given, discounted, or voided entirely. Even credits that have already been dispersed can be voided, invoking a fee on the part of the project manager/credit provider. At the end of the ACL, all remaining credits will be distributed and all Carbon Bonds will mature. Any credits still considered valid at the end of the ACL are considered semi-permanent sequestration. Commonly 10-25 years for Wood Vault projects, but can vary by project and accreditor.

It is recommended that such a scheme be accompanied by an insurance plan as these instruments begin to emerge. As the scientific and engineering understanding of WHS and Wood Vault construction develops and matures, insurance should account for a smaller and smaller portion of costs.

## 5.5 Data archive and information availability

Archive all project reports and associated data and documents for credits electronically, and make it available to carbon market, buyers, and other authorized entities upon request. The data will also be critical for governments and research institutes to assess the efficacy of the method when implemented at large scale worldwide.

# 6  Scientific Basis of Wood Harvesting and Storage

## 6.1  Carbon removal and sequestration to combat climate change

Our atmospheric $CO_2$ concentration has increased from a pre-industrial value of 280 ppm to over 410 ppm today, mostly due to anthropogenic sources, primarily carbon emissions from fossil-fuel burning and deforestation. The Paris Climate Agreement and the Glasgow Climate Pact of the United Nations Framework Convention on Climate Change (UNFCCC) envision limiting global warming to 1.5-2°C warming (IPCC 2018), which would require transformation in energy and economic structure, as well as novel technologies to keep atmospheric $CO_2$ concentration at a safe level.

The primary pathway to reduced greenhouse gas emissions is a transition to "low carbon economies," in which energy efficiency is improved and energy production has a much lower carbon footprint by transforming the energy infrastructure to include more renewable technologies and carbon capture and sequestration. Such a transition, however, is quite difficult to accomplish at the rate required to limit global temperature rise to 2°C—the switch to low-carbon infrastructure is a slow process due to a variety of technological, socioeconomic, and political barriers. Thus, carbon removal and storage, namely removing carbon that is already in the atmosphere and locking it away, could play an important role in the cost-effective stabilization of atmospheric $CO_2$ at acceptable levels. Negative emissions will also be needed considering the long lifetime of atmospheric $CO_2$ even after emissions are completely stopped. Indeed, nearly all future emissions scenarios that involve policy intervention assume significant contribution from carbon sequestration.

The removal of $CO_2$ from the atmosphere can utilize physical, chemical or biological methods (National Academies of Sciences Engineering and Medicine (NASEM) 2019) Biological carbon sequestration, hereafter bio-sequestration, relies on plant photosynthesis to capture $CO_2$ and assimilate the carbon into biomass:

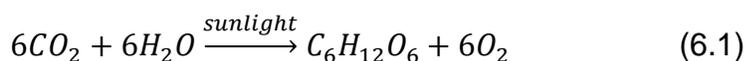

$$6CO_2 + 6H_2O \xrightarrow{sunlight} C_6H_{12}O_6 + 6O_2 \qquad (6.1)$$

The product of photosynthesis sugar (simplest form glucose: $C_6H_{12}O_6$) is further processed into other organic molecules. Respiration is the reverse of the above reaction (Eq. 6.1).

Examples of bio-sequestration include reforestation, no-till agriculture, and intensive forest management. Afforestation or reforestation is arguably the most widely embraced carbon sequestration technique because of its relatively low cost, benign nature and many co-benefits. Unfortunately, its capacity is limited by the availability of land, and the declining rate of sequestration as the forest matures. Because fossil fuel emissions from energy production continue to increase beyond the sequestration capacity of terrestrial ecosystems, mitigation through land-use management is usually viewed as a low-cost approach with relatively modest total mitigation potential.

## 6.2 Wood Harvesting and Storage: Transferring carbon from the fast to slow carbon cycle

The greatest potential for bio-sequestration may not come from one-time carbon storage in live biomass, but from using plants as a 'carbon scrubber' or 'carbon pump'. Notwithstanding the merit of reforestation, the carbon sink diminishes as a forest matures. An alternative is to manage a forest as a 'carbon farm' in a way to separate 'carbon removal' via photosynthesis from 'carbon storage' (Zeng 2008). A fraction of the large biospheric productivity can be siphoned and stored semi-permanently, thus creating a continuous stream of carbon sink.

Net Primary Productivity (NPP) of the terrestrial biosphere is 220 $GtCO_2$ $y^{-1}$, of which woody biomass NPP is 70 $GtCO_2$ $y^{-1}$ (one third of total), and a sustainable harvesting potential for WHS is up to 10 $GtCO_2$ $y^{-1}$ (Zeng, et al. 2013) (Less than 5% of NPP) with which we can absorb about 1/3 of current fossil fuel $CO_2$ emissions rate of 37 GtC $y^{-1}$ (Fig. 6.1). The 10 $GtCO_2$ $y^{-1}$ 'practical' potential has excluded current land use, wood use and conservation needs.

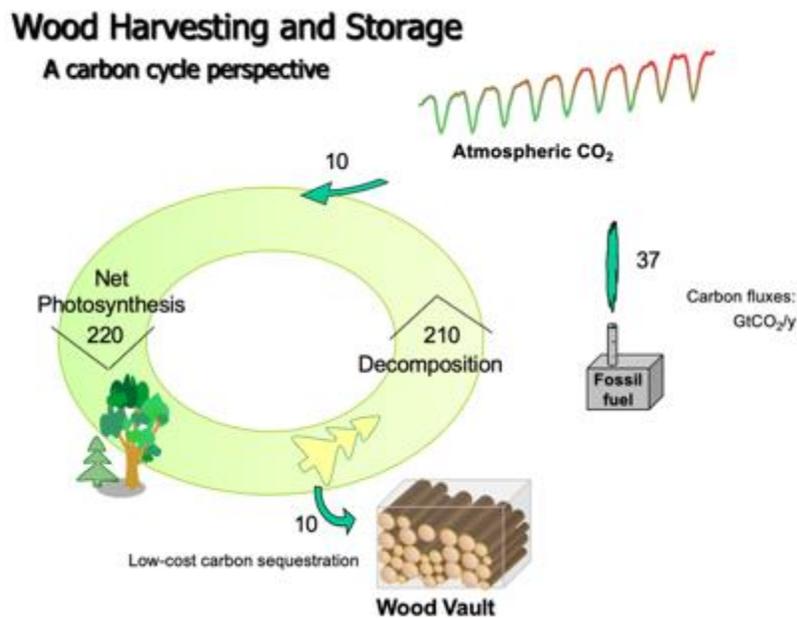

*Figure 6.1. Wood Harvesting and Storage (WHS) siphons off a sustainable fraction of the biosphere production in the form of harvested wood and stores it in engineered Wood Vaults to prevent decomposition, forming an effective carbon sink. WHS transfers the carbon from the 'fast' live biosphere carbon cycle to the 'slow' geological carbon cycle. A WHS carbon*

*sequestration rate of 10 GtCO$_2$ y$^{-1}$ is about 5% of terrestrial net primary productivity (NPP) and 30% of coarse woody biomass (CWB) production.*

Compared to many traditional carbon management ideas in which the stored carbon saturates after a period of time, WHS creates a continuous stream of sequestered carbon. Carbon in stored wood is relatively easy to monitor and verify, reducing risk of loss and other issues facing some other carbon sequestration strategies. In practice, this carbon will come from a variety of sources such as waste wood and managed timber land. The methods of sourcing wood will determine the achievable quantity of net sequestered carbon. The preservation of wood for hundreds of years or longer for climate mitigation exceeds the usual timescales assumed of bio-sequestration such as reforestation or soil conservation. Active management to extend the lifetime of stored wood will be critical for the success. Carbon monitoring and full carbon accounting from source to storage will be essential for carbon credits.

The fact that $CO_2$ capture via natural photosynthesis is 'free', an evolutionary wonder from an ecological (and economic) perspective, puts this process outside the engineering system boundary. This is a great advantage of WHS relative to purely engineered methods such as Direct Air Capture where $CO_2$ capture accounts for the lion's share of cost and energy input. Of course, this advantage is shared by all nature-based methods, thus their popularity. However, the main shortcoming of most nature-based methods is that the permanence of such carbon sinks is generally too short compared to the climate change time scale of hundreds of years. This shortcoming is overcome in WHS by burying wood underground or by other means to ensure highly durable semi-permanent preservation. This requires a dedicated facility carefully engineered to prevent decomposition and an operational process flow that is practical, efficient, and low cost. In summary, WHS is a unique hybrid method that allows human engineers to take advantage of the natural biological process of photosynthesis.

## 6.3   Principles of Wood Vault: How to prevent wood decomposition?

Wood consists mainly of three key components: cellulose, hemicellulose and lignin, with small amounts of extractives (sap etc.) and minerals (ash). Cellulose is a quasi-crystalline polymer formed by thousands to tens of thousands of glucose, which form long fibrous bundles connected and strengthened by shorter hemicellulose and amorphous lignin. While cellulose consists of long chains of glucose, hemicellulose is much shorter, often linking the long cellulose fibrils in a cross-sectional manner with respect to the cell wall, and consists of various sugar building blocks. Lignin is a high molecular weight amorphous polymer of aromatic compounds, contributing to its high resistance to decomposition. Lignin is a main component of soil humus. The typical C:N ratio of wood is 200:1, compared to 20:1 for leaves. Thus, when wood is buried, nutrient loss is relatively small. Importantly, in whole wood structure, lignin forms a scaffolding that protects cellulose and hemicellulose, a critical factor in preventing bacteria attack under anaerobic conditions.

In its natural aboveground setting, woody biomass can be degraded by physical, chemical and biological processes. When buried underground, the physical disturbance will be nearly non-existent except under special circumstances such as earthquakes and landslides. Chemical degradation can occur, particularly in wet environments and especially when in highly acidic conditions, but most natural soil environments are close to pH neutral. This leaves mainly biological degradation of concern.

Biological agents that decompose wood are mainly fungi (brown, white and soft rots), insects (termites, beetles) and bacteria. However, once buried in an anaerobic condition, the majority of these decomposers cannot survive because of lack of oxygen (Blanchette, et al. 1990). This leaves only a small number of anaerobic bacteria that can potentially attack organic matter.

Methane generation can occur in partially anaerobic conditions of a landfill (Wang and Barlaz 2016). Under such conditions, three types of macromolecules, namely proteins, carbohydrates, and lipids provide the main substrates for decomposition of organic matter. The first step is hydrolysis, which creates amino acids, sugar, fatty acids, respectively. Subsequent steps are acidogenesis, acetogenesis and methanogenesis. A suite of bacteria, each of them requiring particular environmental conditions such as suitable temperature ranges, are needed to degrade the organic matter.

However, in an engineered Wood Vault, where only clean woody biomass (no food scraps, toxic chemicals etc.) is buried, methane generation is expected to be negligible for the following reasons. In comparison to material composed primarily of carbohydrate/protein/fat, lignin is difficult to digest for anaerobic bacteria due to its complex polymer structure (Colberg, 1988; O'Dwyer, Walshe, and Byrne 2018). For example, in biodigesters, practically all waste organic material can be digested except for wood. Indeed, a major difficulty in cellulosic ethanol production using woody raw material is how to remove lignin. Often, wood has to be pretreated by physical (heating) or chemical (acid) methods before fermentation (Sayara and Sanchez 2019). Even though resistance of wood to decomposition is an obstacle for bioenergy, here it is a key for preserving wood for carbon sequestration. Another factor that hinders bacteria attack is the extremely low nutrient content in wood as the main lignocellulose structure contains only C, H and O which does not support thriving bacterial communities. Additionally, the wholeness of the buried coarse woody biomass further extends its durability because the physical protection slows down the invasion of already slow (if any) bacteria attack.

Beyond the scientific process understanding, direct evidence of long-term wood preservation comes from landfill studies, archaeological, and geological evidence (Ximenes, Cowie, and Barlaz 2018; Bjordal 2012; Mustoe 2018). In a striking example, wood stumps from a 2.5 million years old ancient forest are well preserved in a clay quarry (Mustoe 2018), an environment similar to Wood Vault engineering design.

While creating anaerobic conditions is one pathway to preserve woody biomass for carbon sequestration, storing woody biomass in dry or cold conditions can also preserve woody biomass semi-permanently because of the fundamental biology that decomposers also need moisture and suitable temperature to thrive. The control of these 3 factors: oxygen, moisture, and temperature, provides the basis for methods in wood preservation on timescales long enough to contribute to removing atmosphere carbon dioxide for climate change mitigation (Fig. 6.2).

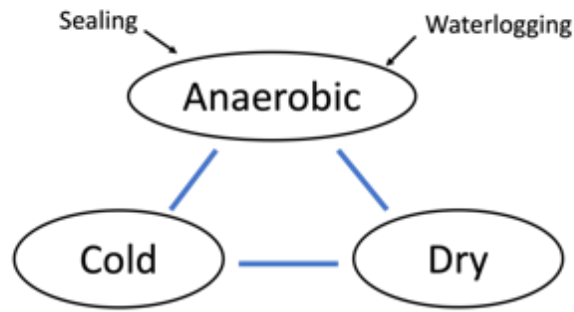

*Figure 6.2. The three main pathways to prevent wood decomposition. Ensuring one or more of the three factors with high fidelity will be sufficient to preserve wood for long timescales relevant to climate change. Source: Zeng and Hausmann (2022).*

# 7  Acronyms

ACL: Active Credit Lifetime
BiCRS: Biomass Carbon Removal and Storage
CRS: Carbon Removal and Sequestration or Carbon Removal and Storage
CDR: Carbon Dioxide Removal
CWB: Coarse Woody Biomass
FWB: Fine Woody Biomass
GHG: Greenhouse Gasses
MRV: Monitoring, Reporting and Verification
NCS: Net Carbon Sequestration
NWW: Natural Waste Wood
UWW: Urban Waste Wood
WHS: Wood Harvesting and Storage
WR: Wood Residual
WV: Wood Vault

# 8 Glossary

**Active Credit Lifetime (ACL):** The period for which the project is being actively monitored and giving out carbon credits. During this period carbon credits can be given, discounted, or voided entirely. Even credits that have already been dispersed can be voided, invoking a fee on the part of the project manager/credit provider. At the end of the ACL, all remaining credits will be distributed and all Carbon Bonds will mature. Any credits still considered valid at the end of the ACL are considered semi-permanent sequestration. Commonly 10-25 years for Wood Vault projects, but can vary by project and accreditor.

**Burial Chamber:** The anaerobic/cold/dry location where sequestered woody biomass is stored. In most common Wood Vault projects this is a subterranean or partially subterranean chamber lined with a low permeability material. It can also be open or closed in cold, dry or water saturated conditions.

**Carbon Bond:** A type of carbon credit issuance scheme where the carbon credit is only distributed at the end of the ACL when the bond matures, assuming certain prerequisites have been met. Commonly used in managed forest sourced projects to ensure that forests are replanted and not clear cut.

**CO2e**: $CO_2$ equivalent of carbon stored in wood.

**Conservation Easement:** A legal framework for the long-term preservation and conservation of a plot of land. It places restrictions on what activities can be done on the plot of land, including restricting it to natural state, agricultural use, or sustainable forestry. It also appoints an entity to conduct periodic monitoring to ensure usage of the plot stays within the limits of the easement. A conservation easement is maintained even after ownership of the land is transferred.

**Dry Biomass:** Biomass that has been dried and reduced to just its structural and other non-water components. Experimentally, it is measured as the mass after wood sample being dried in a 100 degree C oven for 24 hours, or otherwise no further mass loss.

**Net Carbon Sequestration (NCS):** Carbon sequestered semi-permanently in storage after subtracting lifetime loss and $CO_2$ emissions from operation, carbon loss from land use and other impacts.

**Project Manager:** Whatever entity is running and implementing the Wood Vault project. In general, this is the entity implementing or managing the implementation of the steps in this methodology.

**Wet/Green Biomass:** Biomass freshly harvested from its natural state containing significant amount of liquid water.

**Wood Harvesting and Storage (WHS)**: also known as Wood Harvesting and Sequestration. A method to sustainably source woody biomass and store semi-permanently for carbon sequestration.

**Wood Vault (WV):** An engineered structure designed to hold woody biomass in an anaerobic, dry, or frigid environment to prevent decomposition.

**Coarse Woody Biomass (CWB):** Carbon containing biomass with a wood structure of large sizes such as tree stem, branch, or cut logs, with a minimum diameter of 10 cm, as opposed to smaller pieces such as twigs, wood chips and sawdust ("Fine Woody Biomass" [FWB]).

**Wood Residual (WR)**: Residual of a variety of conventional wood uses and wood processing, such as urban waste wood (yard waste), construction and demolition debris, wood from land clearing, wood from forest thinning, often with little or no reuse value.